\documentstyle[times,pre,aps,epsfig,amssymb,balanced]{revtex}
\begin{document}

\title{Dynamics of localized structures in vectorial waves}

\author{Emilio Hern\'andez-Garc\'{\i}a, Miguel Hoyuelos
\cite{presentad}, Pere Colet and Maxi San Miguel}

\address{Instituto Mediterr\'aneo de Estudios Avanzados, IMEDEA
\cite{www} (CSIC-UIB),\\ Campus Universitat Illes Balears, E-07071
Palma de Mallorca, Spain.}

\date{July 12, 1999}

\maketitle

\begin{abstract}
Dynamical properties of topological defects in a twodimensional
complex vector field are considered. These objects naturally arise
in the study of polarized transverse light waves. Dynamics is
modeled by a Vector Complex Ginzburg-Landau Equation with
parameter values appropriate for linearly polarized laser
emission. Creation and annihilation processes, and
selforganization of defects in lattice structures, are described.
We find ``glassy" configurations dominated by vectorial defects
and a melting process associated to topological-charge unbinding.
\end{abstract}


\begin{twocolumns}

A variety of nonlinearly evolving fields display states consisting
of distinct localized objects with some kind of particle-like
behavior. Examples are vortices in fluids, superfluids and
superconductors, solitary waves in chemical media, and oscillons
in granular layers, among others\cite{localized}. These localized
structures often organize the geometry and the dynamics of the
host medium, so that they become `building blocks' of regular
patterns
and of spatiotemporal chaos\cite{vanHecke,toni}. In this
situation, an understanding of complex evolving configurations can
be achieved in terms of the interaction rules of the particle-like
entities. An important class of localized objects are {\sl
defects} that appear as a consequence of spontaneous symmetry
breaking in the surrounding medium. These objects carry
topological properties which endorse them with a characteristic
stability and robustness. Its presence is ubiquitous both in and
out of equilibrium, and often mediate nonequilibrium dynamical
processes.

Optical cavities containing nonlinear optical materials have been
specially prolific in providing examples of localized structures
\cite{scheuer}. They can take the form of vortices or bright or
dark dissipative spatial solitons. These particle-like objects
have been observed or predicted in lasers and photorefractive
materials \cite{photorefractive}, lasers with saturable absorbers
\cite{Rosanov96}, semiconductors \cite{semiconductors} and other
nonlinear optical systems such as saturable absorptive media,
second-harmonic generation, and optical parametric oscillators
\cite{optical}. The ability of `writing', `erasing', and moving
around these localized light spots opens promising ways of
achieving parallel information processing. There is however an
important property of light that is just becoming to be
appreciated in this context: the vector nature of the
electromagnetic field. Sometimes, the polarization degree of
freedom is fixed by material anisotropies or by experimental
arrangements. But, when free to manifest, it leads to striking
topological phenomena \cite{physrep}.

A classification of topological singularities in electromagnetic
fields propagating paraxially in linear media can be found in
\cite{hajnal}. Nonlinear generation and propagation, as occurring
in lasers, favors particular polarization states (circular or
linear, for instance) which should be taken as the relevant
background states on which topological defects may appear in
nonlinear systems. Classification of topological defects on such
vector nonlinear backgrounds was established in \cite{pismen},
improving on earlier work \cite{gil}. While in particular limits
some information on isolated defects may be obtained analytically
\cite{pismen}, there is a lack of understanding about complex
dynamic states. Here we address these unexplored dynamical
properties of the defects for general non-conservative and
non-relaxational dynamics. We study which kinds of defects
spontaneously emerge from the dynamics, and describe their
stability, creation, annihilation and selforganization processes.
Transitions between different regimes of spatiotemporal dynamics,
mediated by defect-behavior changes, are found: A kind of
`vectorial defect' can entrain the whole system and dominate
`frozen' or `glass-like' field configurations whereas
topological-charge unbinding leads to the melting of this frozen
phase into a new dynamical regime.

Our study of defect dynamics is made in the context of a spatially
twodimensional model appropriate for laser emission from
wide-aperture resonators, the Vector Complex Ginzburg-Landau
Equation (VCGLE)\cite{SanMiguel95}. The VCGLE can describe also,
in appropriate ranges of parameters, other kinds of systems such
as two-component Bose condensates\cite{bose}, and
counterpropagating waves in nonlinear
media\cite{counterpropagation}. In general, the VCGLE describes
the complex envelope of any oscillating vector field close enough
to a homogeneous Hopf bifurcation, which leads to a variety of
complex spatio-temporal phenomena. Through this Letter however we
will restrict the study to parameter ranges of interest in optics.

The VCGLE can be written as
\begin{eqnarray}
\partial_t A_\pm = A_\pm &+& (1 + i\alpha) \nabla^2 A_\pm  \nonumber \\
&-& (1 + i\beta) (|A_\pm|^2 + \gamma |A_\mp|^2) A_\pm.
    \label{vcgle}
\end{eqnarray}
$A_\pm$ are the two components of the vector complex field. In
optics they are identified with the right and left circularly
polarized components of the transverse field. Other forms of this
equation can be written \cite{gil} in terms of the cartesian
components $(A_x,A_y)$, which are related to the circular ones by
$A_x=(A_+ +A_-)/\sqrt{2} $ and $A_y=(A_+ -A_-)/i\sqrt{2}$. When
interpreted as a set of two coupled fields the model gives the
opportunity to explore synchronization of spatiotemporal chaos
\cite{toni,hernandez}. The real parameters $\alpha$ and $\beta$
are associated to the strength of nondisipative spatial coupling
(optical diffraction) and nonlinear frequency-shift (optical
detuning) respectively. As for laser systems the condition
$1+\alpha\beta> 0$ is always satisfied \cite{SanMiguel95}, we will
consider only this case. This corresponds to the Benjamin-Feir
stable range, where there are stable plane-wave solutions of Eq.
(\ref{vcgle}). The parameter $\gamma$ (a real number in lasers)
represents the coupling between the polarization components. We
consider a weak coupling situation ($\gamma<1$) so that stable
uniform solutions satisfy $|A_+|=|A_-|$. This corresponds to a
laser emitting linearly polarized light.

For $\gamma=0$ Eq. (\ref{vcgle}) becomes a pair of independent
scalar Complex Ginzburg-Landau Equations (CGLE). A localized
structure which appears in most characteristic configurations of
the different phases of the twodimensional CGLE \cite{chate} is a
topological defect: a ``vortex'' where the amplitude is exactly
zero and thus the phase of the field is not defined. In the regime
where plane waves are stable, a spiral wave develops around the
defect. It behaves asymptotically as a travelling wave (TW) of a
particular wavenumber dynamically selected by the presence of the
defect \cite{hagan}.

When $\gamma \neq 0$ the two components $A_+$ and $A_-$ become
coupled and we have genuine vector effects. The classification of
defects of the vector field in \cite{pismen} was elaborated in
terms of the cartesian components of the field ($A_x$, $A_y$), but
it is better for our purposes to recast it in terms of the
circular components $A_{\pm}=|A_\pm| e^{i\phi_\pm}$: Eq.
(\ref{vcgle}) admits a continuous family of TW solutions for $A_+$
and $A_-$ which are the obvious generalization to $d=2$ of the
onedimensional solutions described in \cite{SanMiguel95}. Defects
appear when different solutions of this family are selected in
different regions of the space. The different solutions can be
matched continuously except at one point, the defect, where the
field has to take a value outside the family. In our case
localized zeroes in $|A_\pm|$ are defects. Two topological charges
$n_\pm$ associated to each defect are defined by
\begin{equation}
n_\pm \equiv \frac{1}{2 \pi}\oint_{\Gamma} d\vec r \cdot \vec
\nabla \phi_\pm \ \ ,
\label{tcharge}
\end{equation}
where $\Gamma$ is a closed path around the defect. We find the
following defects in our dynamical system: A {\it vectorial
defect} is a zero in both components of the field at the same
point. It is of {\it argument} type when $n_+ = n_-$. The
background solutions matched around it are TW with the same
wavevector for $A_{\pm}$. A simple {\sl ansatz} for this solution
reveals that it selects asymptotically the same wavenumber as for
$\gamma=0$. If $n_+ = - n_-$ the vectorial defect is of {\it
director} type and the background solutions are TW for $A_{\pm}$,
again with the same asymptotic wavenumber, but with different
wavevector orientation. Finally, we call {\it scalar defect} a
zero in just one of the two circular field components. The
background solutions are TW. The wavenumber in the component
containing the zero decreases with $\gamma$, being the one in the
other component always vanishing.
Numerically, we do not find
defects with $|n_\pm|>1$. A variety of defect interaction
processes are possible which respect the necessary requirement of
topological-charge conservation. We now describe some of these
processes that occur in our dynamical model.

We first consider\cite{numerics} the spontaneous formation of
defects starting from random initial conditions around the
unstable solution $A_+=A_-=0$. For short times the dynamics
creates a high density of scalar defects. Those of opposite charge
in the same component of the field may collide and annihilate in
pairs during this transient. At a later stage, and for $\gamma$
not too large, vectorial defects are formed due to the coalescence
of two scalar defects, belonging to different field components,
which form a bound structure. This later stage is reached later as
$\gamma$ becomes smaller. Of course, for $\gamma=0$ the lack of
coupling between the components precludes the formation of
vectorial defects. Close to the potential limit $\alpha=\beta$
\cite{potential} (which includes the real-coefficient case
$\alpha=\beta=0$) they are neither formed. The appearance of
vectorial defects has a strong influence on the dynamics: spiral
waves develop around the core in each component which immediately
expel out all the scalar defects. Thus the vectorial defects
become the organizing centers of the final field configurations
(see Fig.~\ref{campos}). The rotation sense of the spiral in each
component is determined by the sign of the corresponding
topological charge (see figure caption). The long-time
configurations are characterized by a structure of cellular
domains with nearly constant modulus separated by shocks between
the waves. These `frozen' or `glassy' configurations with an
extremely slow evolution of the modulus of the field components
look similar to configurations found in the scalar CGLE. However,
these configurations arise here from the dynamical dominance of
the vectorial defects, behaving the scalar ones more passively. In
the scalar CGLE the difference between the dominant defects and
the ones at the domain borders seems to arise from spontaneous
amplification of inhomogeneities\cite{aranson2,hendrey}. Here only
the scalar defects accumulate at the domain borders. The
polarization state in the domain around an argument defect is one
of constant linear polarization, with direction determined by the
phase difference between the spirals. The state around a director
defect is also linearly polarized, but with polarization direction
rotating around the defect core. Scalar defects do not present a
developed spiral wave around them in the dynamical situation just
described, although it may appear in the charged component when
initial conditions producing well-separated scalar defects are
used. Its core is circularly polarized.

The frozen configurations occur for relatively small $\gamma$.
When increasing $\gamma$ vectorial defects become dynamically
unstable and they are destroyed leading to a `melting' of the
glass phase. We have identified two mechanisms in which this
process takes place: (i) Background instability: one of the two
charges forming a vectorial defect is annihilated by an external
scalar defect; as a result a free scalar defect is left in the
other component of the field. (ii) Core instability: the vectorial
defect splits in two scalar defects.

The region in parameter space $\alpha$-$\beta$ where process (i)
is observed corresponds approximately to the region where for the
scalar CGLE the phase spirals are convectively unstable
\cite{aranson}. The spirals remain in place and look stable
because perturbations are effectively ejected away thanks to the
group velocity on the spiral wave. As $\gamma$ is increased from
zero the stability of the spirals is modified: At a given value of
$\gamma$, the group velocity is not strong enough to overwhelm the
growth of the perturbations, the spirals becoming absolutely
unstable. At this point the domains around the vectorial defects
are uneffective as exclusion zones, so that scalar defects
previously confined to the domain border can approach the
vectorial defect core (Fig.~2). This allows for mechanism (i) to
take place. Although this picture is valid for both kinds of
vectorial defects, director defects survive for larger $\gamma$
than argument ones. For the parameter values of Fig.~\ref{campos},
argument defects become unstable at $\gamma
\simeq 0.3$, while director
defects remain stable up to $\gamma \simeq 0.35$. For larger
$\gamma$ only scalar defects are found numerically. The different
stability range of argument and director defects can be understood
through a linear stability analysis of the vector spirals focusing
in its far-field plane-wave structure. An extension of the $d=1$
analysis \cite{SanMiguel95} indicates that as $\gamma$ is
increased, polarization phase instabilities are such that
co-rotating spirals, corresponding here to background solutions
for argument defects, become {\sl convectively} unstable before
than the counterrotaing spirals associated with director defects.
The calculation of the absolute instability limit is quite
involved, but the result for the convective instability suggests
that co-rotating spirals become absolutely unstable before than
counterrotating spirals.

Process (ii) is roughly present in the region of parameter space
$\alpha$-$\beta$ where the spirals are stable in the scalar case.
The splitting of a director defect is shown in
Fig.~\ref{split_dir}. The size of the vectorial defect-core is
much smaller than the size of the core of the two scalar defects
that remain at the end of the process. Also in this case argument
defects become unstable for smaller $\gamma$ than director
defects. For example, for the parameter values of
Fig.~\ref{split_dir} argument defects already split for $\gamma =
0.75$. The splitting mechanism has been previously described in
Ref. \cite{pismen} for the real-coefficient case
($\alpha=\beta=0$), where a greater symmetry between director and
argument defects seems to be present. More in general, approaching
the line $\alpha=\beta$, we observe numerically that director and
argument defects of initial configurations such as the one in
Fig.~\ref{campos} split spontaneously, even for very small values
of $\gamma$.

For $\gamma$ high enough, the vectorial defects always disappear
following one of the two mechanisms described above. The system
then presents a faster
disordered dynamics (a kind of {\em gas-like} phase) dominated by
the scalar defects, which are conserved in number during very long
times. A typical snapshot is shown in Fig.~\ref{anticor2}. At the
defect core one of the component vanishes, and the modulus of the
other has a local maximum. Thus the localized objects are
circularly polarized and impose some elliptic polarization to
their neighborhood. The spiral wavelength around scalar defects
increases with $\gamma$, so that well-developed spirals do not fit
in the domains for $\gamma$ close to one. Domains are thus less
effective as exclusion zones and defects more mobile.
The
vortex unbinding transition between the glassy and the gas-like
phases can be described quantitatively in terms of entropy and
mutual information measures \cite{hernandez,hoyuelos}.

We finally discuss the emergence of self-organized ordered
structures of defects. In the dynamics leading to configurations
as the one in Fig.~\ref{campos} there are cases, particularly for
small $\gamma$, in which only one or few vectorial defects are
formed. They immediately push the scalar defects out of the limits
of their large domains, so that a large number of scalar defects
are compressed in a limited region of space. In this situation the
`gas' of scalar defects `crystallizes' forming a stable square
lattice with alternating positive and negative charges as in an
ionic crystal (see Fig.~\ref{cristal}). The lattice in one of the
components fills the interstitials of the other. Once the lattice
is formed, the vectorial character of the field is no longer
required to keep the lattice stable. In fact, crystalline
aggregation of defects was previously observed in the scalar CGLE
in special situations (when $\alpha$ is close to $\beta$, see
\cite{aranson2}). Note, however, that in the scalar case very
special initial conditions are needed to obtain the lattice of
defects whereas here the vectorial defect creates a large
exclusion island which compresses the scalar defects and leads to
the spontaneous condensation, in rather general conditions, of a
highly dense cristal.

In summary, we have described dynamic phenomena associated with
defect dynamics in vector nonlinear media. Crystal-, gas-, and
glass-like phases are found with transitions between them mediated
by processes in which the vector nature of the field and the
defects plays an important role. Optical active media are the
natural systems in which to search for experimental realizations
of these phenomena. In addition to applications to information
storage and processing, the particle-like objects studied here
present the concurrent appearance of spatial localization both of
light intensity and of optical polarization. This makes them very
interesting from the point of view of applications leading to atom
trapping and cooling.

Financial support from DGICYT (Spain, Projects PB94-1167 and
PB97-0141-C02-02) and from the European Commission (TMR Project
QSTRUCT FMRX-CT96-0077) is acknowledged. M. H. acknowledges also
support from the FOMEC project 290, Dept. de F\'{\i}sica FCEyN,
UNMP, and from CONICET grant PIP No. 4342.


\begin{figure}
\psfig{figure=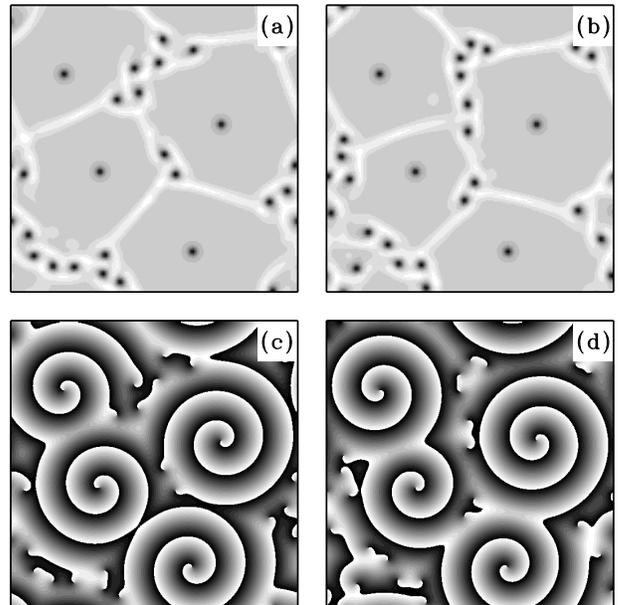,width=8cm} \vspace{0.1cm}
\caption{Long-time field
configurations for $\gamma = 0.1$, $\alpha=0.2$ and $\beta=2$. (a)
$|A_+|^2$, (b) $|A_-|^2$, (c) $\phi_+$, and (d) $\phi_-$. In (c)
and (d), the upper-left and lower-right spirals have the same
sense, thus identifying its core as an argument defect;  the other
spirals wind in opposite senses (director defects).}
\label{campos}
\end{figure}

\begin{figure}
\psfig{figure=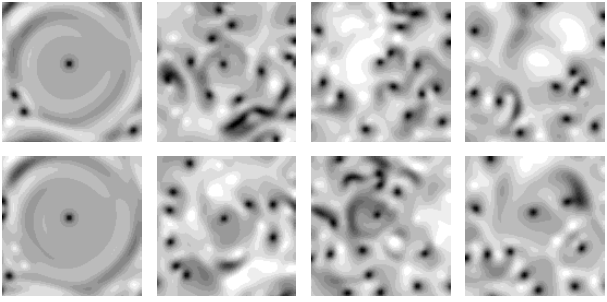,width=8cm} \vspace{0.1cm}
\caption{Annihilation of a director defect ($\gamma = 0.35$,
$\alpha=0.2$, $\beta=2$). Upper row: $|A_+|^2$, lower row:
$|A_-|^2$. From left to right: $t=90$,$190$,$270$,$290$. The
initial condition was the configuration of Fig.~\ref{campos}. Only
a part of the simulation domain is shown. }
\label{aniq_dir}
\end{figure}

\begin{figure}
\psfig{figure=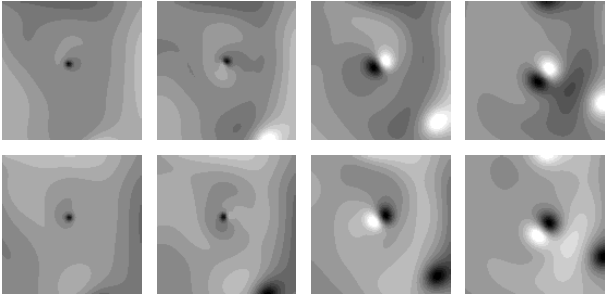,width=8cm} \vspace{0.1cm}
\caption{Splitting of a director defect ($\alpha
=0.7$, $\beta=2$, $\gamma=0.95$). Upper row: $|A_+|^2$, lower row:
$|A_-|^2$. From left to right: $t=50$,$100$,$150$,$200$. The
initial state formed spontaneously under $\gamma=0.9$. }
\label{split_dir}
\end{figure}

\begin{figure}
\psfig{figure=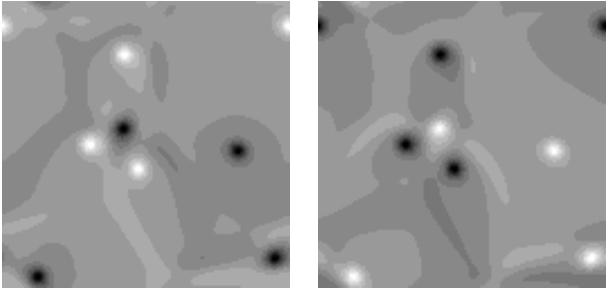,width=8cm} \vspace{0.1cm}
\caption{An evolving gas-like state, dominated by scalar
defects ($\alpha=0.2$, $\beta=2$, $\gamma =
0.8$). Left: $|A_+|^2$, right: $|A_-|^2$.}
\label{anticor2}
\end{figure}

\begin{figure}
\psfig{figure=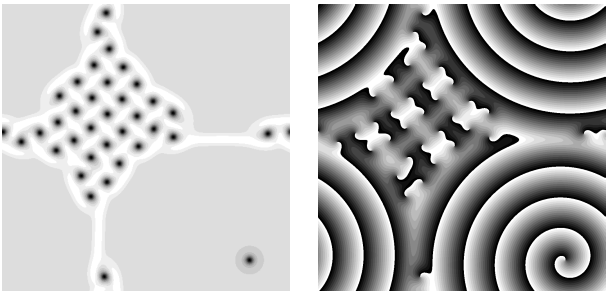,width=8cm} \vspace{0.1cm}
\caption{Crystal of defects, compressed by the vectorial defect in
the lower-right corner ($\alpha =0.2$, $\beta=2$, $\gamma=.01$).
Condensation occurred spontaneously starting from random initial
conditions. Left: $|A_+|^2$; right: $\phi_+$. }
\label{cristal}
\end{figure}


\end{twocolumns}


\begin{thebibliography}{99}
\bibitem{numerics} We use a numerical
pseudospectral method with second order accuracy in time
\cite{hoyuelos} in lattices of size $128 \times 128$. The spatial
and temporal discretizations were $\Delta x=1$, and $\Delta t=
0.05$.
\end{thebibliography}
\end{document}